\RequirePackage{fix-cm}
\RequirePackage{amsmath}

\documentclass[12pt]{article}

\usepackage{graphicx}
\usepackage{amssymb}
\usepackage{verbatim}
\usepackage{natbib}
\usepackage{bm}
\usepackage{etoolbox}

\begin{document}

\title{\bf Bootstrapping Through Discrete Convolutional Methods}

\author{
Jared M. Clark \\
and \\
Richard L. Warr \\ \\ Department of Statistics \\ Brigham Young University \\ Provo, UT 84602, USA
}

\maketitle
\thispagestyle{empty}

\begin{abstract}
Bootstrapping was designed to randomly resample data from a fixed sample using Monte Carlo techniques. However, the original sample itself defines a discrete distribution.  Convolutional methods are well suited for discrete distributions, and we show the advantages of utilizing these techniques for bootstrapping.  The discrete convolutional approach can provide exact numerical solutions for bootstrap quantities, or at least mathematical error bounds.  In contrast, Monte Carlo bootstrap methods can only provide confidence intervals which converge slowly.  Additionally, for some problems the computation time of the convolutional approach can be dramatically less than that of Monte Carlo resampling.  This article provides several examples of bootstrapping using the proposed convolutional technique and compares the results to those of the Monte Carlo bootstrap, and to those of the competing saddlepoint method.
\end{abstract}

\noindent {\it Keywords:}  Convolution, degradation data, discrete Fourier transform, and saddlepoint method

\section{Introduction}
\noindent
Bootstrapping is a resampling technique that relies on taking random samples with replacement from a data set. Bootstrapping techniques \citep{10.2307/2958830, efron1994introduction} provide researchers with an increased capacity to draw statistical inference from a single sample, whether or not parametric assumptions are made. The theoretical bootstrap distribution of a statistic exists, but as sample size (n) increases, the number of possible bootstrap samples grows quickly with $n^n$ possibilities. 
Monte Carlo methods are standard practice for approximating the bootstrap distribution of a statistic.

While the Monte Carlo bootstrap is extremely versatile, as with all procedures, it has some limitations. Monte Carlo bootstrap methods are relatively straightforward to implement but can be computationally intense, and produce stochastic error bounds which narrow at the relatively slow rate of $O(1/\sqrt{N})$. This article introduces an alternative method for bootstrapping quantities using convolutional methods which, when short computation times and high levels of accuracy are valued, often proves to be better suited than the Monte Carlo approach and the competing saddlepoint method. The convolutional method is computationally fast and provides either exact values or more precise approximations.

For a simple illustration, consider the sample mean, $\bar{x}$, from a sample \{$x_i$\} for $i = 1,2,\ldots,n$.  The sample implicitly defines a discrete distribution with the support $\{x_1,x_2,\ldots,x_n\}$ with each atom being equally probable (assuming no ties).  If for instance, one is interested in finding the bootstrap distribution of the mean, it is primarily characterised by self-convolving a modified version of that distribution $n$ times. Since the distribution of the bootstrap mean is defined by the sum of discrete random variables with compact support, it also has a discrete distribution with compact support. This simple scenario outlines the basics of the method described in this article.

The numerical techniques for computing convolutions are numerous, however in this work we focus on using the discrete Fourier transform (DFT). The formulae for computing both the DFT and its inverse have accompanying algorithms in the fast Fourier transform (FFT). These algorithms are readily available on most, if not all, computational platforms.  The FFT scales extremely well 
according to $O\left(N\log N\right)$ complexity \citep{cooley1965algorithm}.
In some instances, the method we propose can compute the bootstrap distribution exactly.  However, in the more general case, when discretization error is introduced, mathematical bounds can be obtained for quantities of interest.  

We provide an outline of the remainder of the paper. First, a review of the relevant literature is presented. Once the literature has been discussed, the methodology is explained. The next section focuses on a comparison between the discrete convolutional and Monte Carlo methods. The comparison is followed by a section containing five data applications of the discrete convolution method. The examples also serve as an additional comparison between computational approaches. Finally, in the conclusion we summarize the main points of this paper.

\subsection{Literature Review}

\noindent
Exploring alternatives for Monte Carlo bootstrapping techniques is not uncommon. There have been a few lines of research that have produced excellent results. In this section we list some of these works and their relevant findings.

The most prominent alternative approach to approximate bootstrap distributions is the use of the saddlepoint method. In those endeavors the saddlepoint approximation inverts the moment generating function of a random variable to obtain an accurate estimate for the probability density function.  For example, \cite{davison1988saddlepoint} developed methods to estimate the bootstrap distribution of a statistic using saddlepoint approximations.  Other work along these lines includes \cite{butler2002bootstrapping}, which bootstrapped survival curves using a saddlepoint approximation. This same paper focused on using the saddlepoint method to estimate the mean and variance of first passage times (FPT). In an example studying the progression of dementia, the saddlepoint method was generally able to predict values closer to the truth than the Monte Carlo bootstrap. 

More recently, the saddlepoint method was used to approximate first passage quantiles of a stochastic process.  For example, \cite{balakrishnan2019nonparametric} developed a nonparametric method to approximate FPT quantiles of a degradation process. This methodology was further improved in \cite{palayangoda2020improved} which increased the accuracy and proposed a method to handle unequally spaced data. Given the popularity of the saddlepoint method, it will be used as one of the main comparisons for the examples in this article and we will compare our proposed methods directly with results from both of these works as well as others. 

The method proposed in this article is based on the discrete Fourier transform and its celebrated computational implementation, the fast Fourier transform. The convolutions inherent in a bootstrap distribution are computed in the Fourier transform space. 
Previous work has applied the FFT to provide exact p-values \citep{beyene2001uses}. FFT methods have been used to calculate the distribution of a random variable, as well as find exact hypothesis test p-values for general linear models. 

Additional research has calculated the distribution of convolutions with the FFT. \cite{kern2003using} used the FFT to approximate a two-dimensional distribution that characterizes animal locations. Although the estimated distributions are different from those approached in this paper, use of the FFT was shown to drastically decrease computation time.  Additionally, \cite{warr2014numerical,warr2020error} provided mathematical bounds on the convolutions of random variables using the FFT.   
We use these established methods to accurately calculate quantities from bootstrap distributions.

Recent research has also developed the algorithmic construction of bootstrap confidence intervals \citep{efron2020automatic}. Although the goal of that research is not to revolutionize the estimation method of a bootstrap distribution, it demonstrates that bootstrapping methods continue to improve and are becoming an ever more viable option for inference. Furthermore, \citeauthor{efron2020automatic} state that the bootstrap confidence interval often enjoys better accuracy over intervals constructed using large-sample normal approximations. 

\section{Methodology} \label{sec:methods}

\noindent
The typical Monte Carlo bootstrap is performed by randomly resampling from the data.  In this section we propose an alternative method that computes the bootstrap distribution directly.

\subsection{Bootstrap Distributions}

To establish notation we define a distribution, $F$, such that we have $n$ mutually independent random variables $X_i \sim F$ for $i \in \{1,2,\ldots,n\}$.  Once the random variables are drawn from their distribution, they are no longer random, and we denote the observed sample as $x_1,x_2,\ldots,x_n$ or just $\bm{x}$. Now a discrete probability measure can be defined from the data such that 
\[ Y \sim G \ \text{, where } G =  \frac{1}{n}\sum_{i=1}^{n} \delta_{x_i}.\]
We define $\delta_{x_i}$ to be the Dirac measure at $x_i$. The sample $\bm{x}$ now defines a discrete distribution. 

The random variable $Y$ itself is typically not of interest, however, we are often interested in statistics which are functions of $Y$.  Thus we denote a generic bootstrap statistic of $Y$ as $T(\bm{y})$.  Our proposed method is intended for cases where the bootstrapped statistic can be written as the sum of independent random variables.

\subsection{Convolutions using the Discrete Fourier Transform}

The sum of independent random variables is an example of a convolution.  Similar to using moment generating functions, the discrete Fourier transform provides a readily available technique to calculate the convolution defined by the sum of independent random variables.  

The discrete Fourier transform is the primary tool in our method to calculate the distributions of bootstrap statistics. To use the DFT, an equally spaced grid of length $N$ is defined on the support of the random variable $Y$.  This grid must be defined such that it adequately covers the support of $Y$ and $T(\bm{y})$. We denote the grid on the support to be $\bm{s} = \{s_0, s_1, ..., s_{N-1}\}$. Then the DFT for $Y$ is the sequence defined by:
\[ \tilde{f}_{Y,k} = \sum_{j=0}^{N-1} P(Y=s_j) \, e^{-\frac{i2\pi}{N}kj} \]
for $k \in \{0,1,\ldots, N-1\}$.

If $Y$ were to be convolved with a second random variable, $Z$, the DFT for $Y+Z$ is the sequence defined by $\tilde{f}_{Y+Z,k} = \tilde{f}_{Y,k} \times \tilde{f}_{Z,k}$, so long as $Y$ and $Z$ are independent. 

In order for this convolutional method to be exact, $\bm{s}$ must precisely align with the supports of $Y$, $Z$ and $Y+Z$. Here we note that the grid $\bm{s}$ and the actual supports of those random variables may not align.  If this is the case, an approximate distribution for $Y+Z$ can be found and the error bounds of the approximation can be obtained.  It is also important to point out that random variables defined by a sample, such as $Y$, have discrete and compact support.  The discreteness property aides in finding an appropriate distance between points in $\bm{s}$, and the compactness property allows for straightforward selection of a starting and end point for $\bm{s}$.

The basic approach for bounding the error of an approximated convolution is: if all exact support points (not contained on the grid $\bm{s}$) of the random variables are rounded down to the next lower grid point, and the DFT convolution method is used, the resulting CDF will be an upper bound for the distribution. Likewise, if all exact support points of the random variables are rounded up to the next greater grid point, the result is a lower bound for the CDF. Using this approach, it is possible to mathematically bound the true convolved distribution in cases where it isn't practical to ensure the support of the random variables and the grid $\bm{s}$ completely agree.

One desirable feature of the DFT is the ease in calculating both the forward and inverse DFT using a fast Fourier transform algorithm. Using the FFT and its inverse, it becomes a simple calculation to find the DFT, convolve in the Fourier domain, and then use the inverse transform to find the CDF of the bootstrap statistic.

The process for finding the bootstrap distribution of a statistic (which is defined by a sum of independent random variables) can be described in a few simple steps.  First, the random variables to be convolved should be identified. These will be some function of the data. For these statistics, the convolved random variables should be independent, but don't necessarily need to be identically distributed.  Next, a grid, $\bm{s}$, should be defined. This grid needs to be large enough to contain the full support of $Y$ and the bootstrap statistic, $T(\bm{y})$. The grid consists of equally spaced points. Note that the number of grid points will determine the accuracy of this method.  In general, more grid points will produce more accuracy.  When the support of the distribution and the grid agree, the resulting distribution will be the exact distribution of the bootstrap statistic. Also note that some FFT algorithms are more efficient if the number of the grid points in the support is a power of two.  Third, the probability mass functions for the random variables to be convolved must be defined. The probability mass functions will need to agree with or be fit to the the support grid, $\bm{s}$. Next, the DFT for each random variable is found (utilizing the FFT).  The DFTs can then be multiplied together to perform the convolution. Finally, by inverting the resultant DFT, the pmf of the bootstrap distribution is obtained.  This process will be demonstrated several times in the examples and applications.

\subsection{The Bootstrap Distribution of the Mean}

The sample mean is one of the most commonly bootstrapped statistics. The distribution of the bootstrap mean can be defined as follows:
\begin{equation} \label{eq:bstpMean}
\bar{Y} = \frac{1}{n} \sum_{i=1}^{n} Y_i, \text{ for } Y_i \ {\buildrel \rm \textit{iid} \over \sim} \ G. 
\end{equation}
The distribution of $\bar{Y}$ provides an estimate for the population mean (with reasonable stochastic uncertainty) when the distribution $F$ has finite variance \citep{knight1989bootstrap}.

One can certainly obtain Monte Carlo samples from the distribution of $\bar{Y}$, but the discreteness of $G$ invites the use of discrete methods to obtain the distribution of $\bar{Y}$, which reflects our uncertainty regarding the true population mean.  This can be done using the convolution defined in Equation \ref{eq:bstpMean} and the discrete Fourier transform.  A consequence of the discrete nature of $G$ is that the distribution of $\bar{Y}$ and various other random quantities also have discrete distributions.   

For this example let $w_i=1/n*(x_i- \min \{\bm{x}\})$. The shifting of $\min \{\bm{x}\}$ ensures that the support is simple to work with. In order to employ the DFT to find the bootstrap mean's distribution, we define an equally spaced grid, $\bm{s}$, of length $N$.  When choosing $\bm{s}$, we attempt to include all the $w_i$ values and define it on the interval $[0,n \times \max \{ \bm{w} \}]$. The $w_i$'s define a probability measure, $H$, on $\bm{s}$ such that each $w_i$ contributes mass of $1/n$.  Many grid points, $s_j$, will have a mass of 0.  Let the random variable $Z \sim H$ and $\tilde{f}_{Z,k}$ denote the DFT sequence of $Z$ for $k \in \{0,1,\ldots,N-1\}$. The DFT sequence of $V=\sum_{i=1}^{n} Z_i$, where the $Z_i$'s are independent draws from $H$, is:
\begin{equation} \label{eq:bstpMean2}
\tilde{f}_{V,k} =  (\tilde{f}_{Z_1,k})^n.
\end{equation}

The pmf for the random variable $V$ is simply found by invoking the inverse discrete Fourier transform on the sequence defined in Equation \ref{eq:bstpMean2}. A common definition of the inverse DFT is:
\begin{equation} \label{eq:inverseDFT} P(V=s_j) = \frac{1}{N} \sum_{k=0}^{N-1} \tilde{f}_{V,k} \, e^{-\frac{i2\pi}{N}kj} \end{equation}
for $j \in \{0,1,\ldots, N-1\}$.  Note, that after employing the inverse FFT, some implementations omit the $1/N$ term. This is certainly the case in \texttt{R} \citep{R}, and the result must be divided by $N$ afterwards. 
Once shifted by $\min \{ \bm{w} \}$ the pmf of $V$ becomes the bootstrap distribution of the mean, or in our more standard notation the distribution of $\bar{Y}$.

The primary issue with employing this convolutional method is that it requires an evenly spaced grid on the support which includes all values $x_i/n$.  This is often not tenable in practice.  however, it is possible to create a grid that nearly contains each value $x_i/n$ and the mass assigned to that value can be reassigned to the nearest grid point.  While some precision will be lost, this process provides a readily obtainable approximation.  We also show how to bound the error of that approximation.

\subsection{Illustrative Examples}

This first example investigates an extremely simple case. However, the example should provide some intuition into how the convolutional method works. 

\begin{figure}
    \includegraphics[width=80mm]{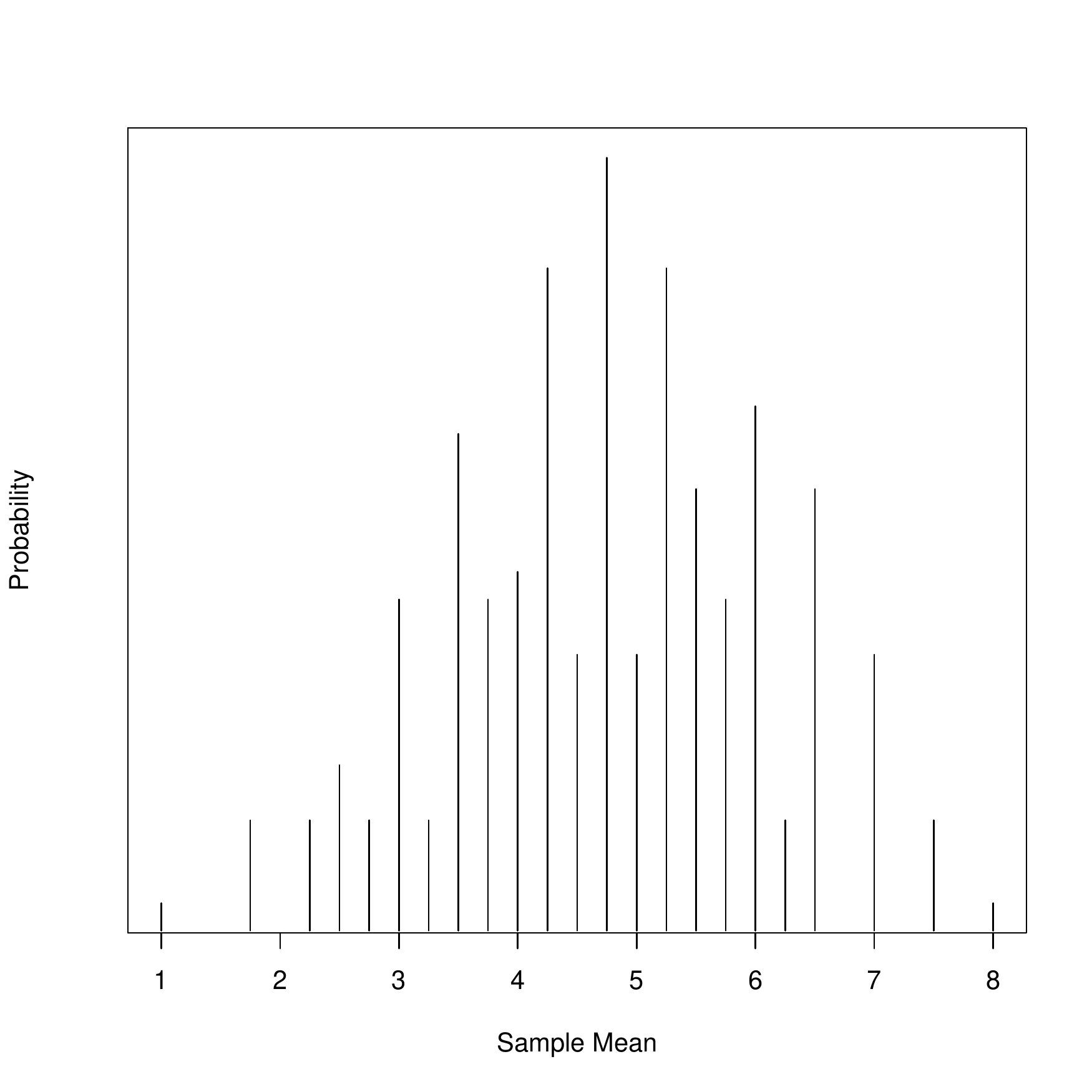}
    \centering
    \caption{In the case of just 4 observations, it is relatively simple to find the exact bootstrap distribution of the sample mean.}
    \label{fig:ex3}
\end{figure}

Consider the sample $\bm{x} = \{1, 4, 6, 8\}$. In this instance, due to the small sample size, it is easy to find the exact bootstrap distribution of $\bar{Y}$ by enumerating the sample space. The exact distribution is displayed in Figure \ref{fig:ex3}.

To apply the discrete convolutional method described in this paper, the support needs to be broken into a grid of equidistant support points. For this example, the grid was defined as the sequence of numbers from 0 to 8 increasing in increments of 0.25. This support was chosen as it includes each $x_i/n$ and covers the entire support of $\bar{Y}$.  The results are exact (up to numerical precision).

After dividing the observations by the sample size of 4 there are now four observed contributions that match with four of the predetermined grid points (0.25, 1.00, 1.50 and 2.00). A probability sequence of the same length as the support should now be generated. Where an observation matches a support point, the value in the new sequence will be 1/4, while all other values will be 0. That is, the term 1/4 should be found in the second, fifth, seventh and ninth positions. This characterizes the pmf of the discrete random variable $Y/n$.

Now properties of the discrete Fourier transform can be used to find the bootstrap distribution of the sample mean. This sequence of probabilities should be transformed using the DFT. Next the sequence will be raised to the $n^{th}$ power, in this case the fourth power, to account for the convolution. This results in the DFT for the bootstrapped sample mean.  Finally, the sequence can be un-transformed, using the inverse FFT, and it provides the exact pmf of $\bar{Y}$.

This is one of the simplest applications of the convolutional method. An additional step in complexity is added when there is no obvious way to create an equally-spaced grid that contains the entire support of the statistic of interest. Now consider changing the previous example by replacing an observation such that $\bm{x} = \{1, \pi, 6, 8\}$.

An intuitive approach would be to create a grid and then round observations to the closest grid points. 
However, we advocate for a slightly different approach, which allows the error of the approximation to be mathematically bounded.  The process for finding an approximation is similar to the exact process shown above, but the process is repeated twice, once for the upper bound and once for the lower bound.  

To create a lower bound on the cumulative distribution function (CDF) of interest, rather than rounding the data, each data point can be shifted up to the nearest higher grid point. Likewise, finding the upper bound shifts the mass of each observation to the nearest lower grid point.

The theory for the process of bounding a distribution through convolutional methods was presented in  \cite{warr2020error}. By computing the difference of the upper and lower bounds of the CDF at any of the grid points on the support, a mathematical bound is placed on the distribution of $\bar{Y}$.

The width of these bounding intervals can be changed by increasing or decreasing the number of grid points. With more support points
the interval widths will be narrower.

Referring back to the example; to create the bounds, the grid was taken as a sequence from 0 to 9 with a difference of 0.1 between adjacent terms. Once the grid was defined, the data were fit to the grid as described above and the convolutional method was employed to place bounds on the distribution.

\begin{figure}
    \includegraphics[width=80mm]{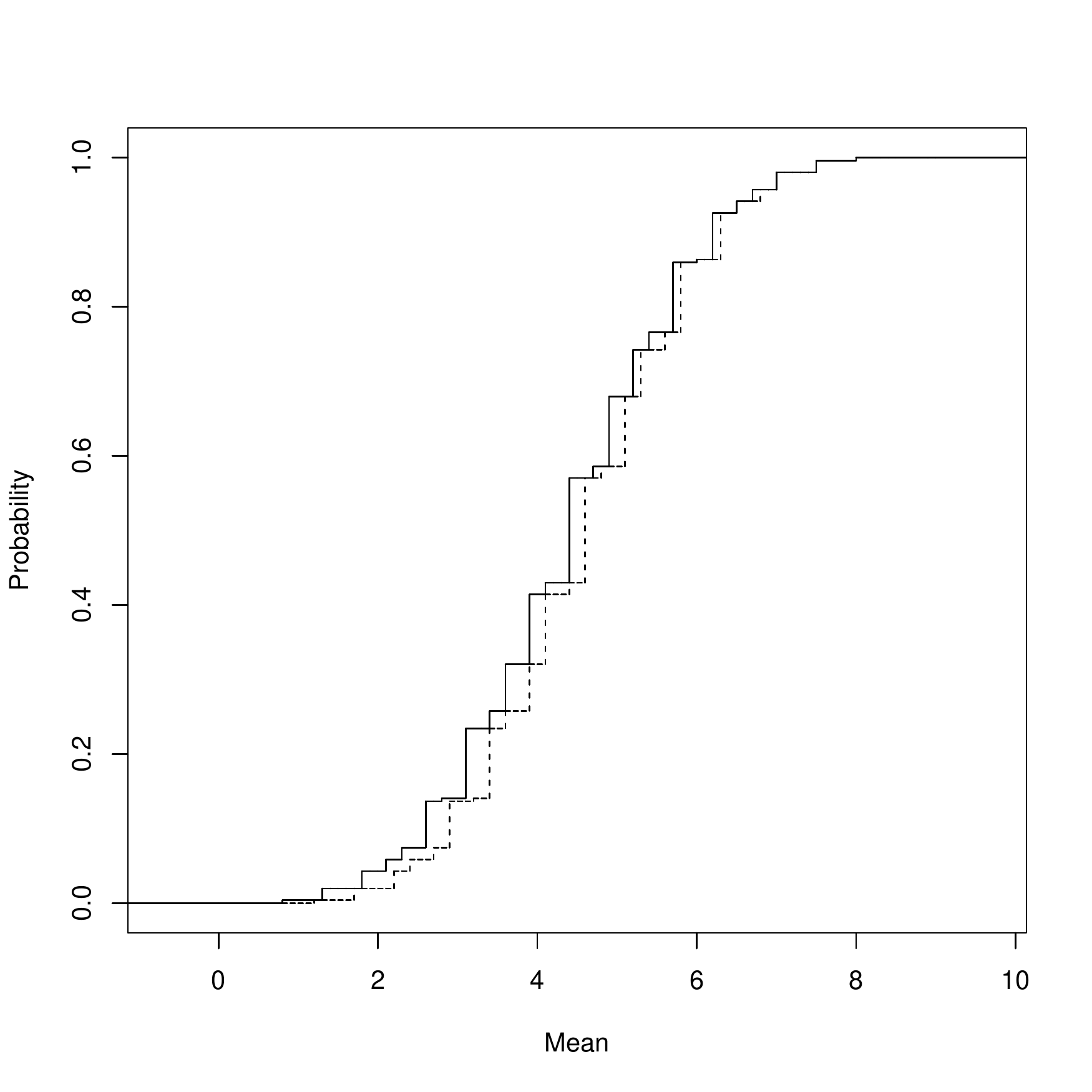}
    \centering
    \caption{The CDF for the bootstrap distribution of the mean can be bounded. The upper bound is given by the solid line, while the dashed line is the lower bound. The true CDF is contained between the upper and lower bounds.}
    \label{fig:bounds}
\end{figure}

As shown in Figure \ref{fig:bounds}, even when it isn't feasible to find the exact bootstrap distribution, the distribution can be bounded. It would be simple enough to increase the number of grid points which would result in a more accurate approximation of $\bar{Y}$'s distribution.

\subsection{Bootstrapping Degradation Data} \label{subsec:MethodDeg}

In addition to the sample mean, the convolutional method is effective in bootstrapping other quantities.  In this section we discuss an approach to find quantiles of the bootstrap distribution of a product's failure time using degradation data. 
Some of the applications presented later in this article demonstrate this in detail.

In the simplest case, consider a collection of items that experience a random amount of degradation $X_i \sim F$, for $i \in \{1,2,\ldots,n\}$, which occurs during a fixed time period $d$.  That is, if the system is inspected in time increments of $d$, the increase in cumulative degradation since the last inspection is drawn from the distribution, $F$. In this case we consider the $X_i$ to be independent and identically distributed random variables.  Once observed, the $x_i$ define a discrete random measure and we assign the random variable $Y$ to that distribution.  If the total amount of degradation meets or exceeds some predefined threshold, $T>0$, then we consider an item to have failed.  

Thus when using Monte Carlo sampling to bootstrap degradation data for failure times, we sample independent $Y_i$ until their sum meets or exceeds $T$.  Suppose we sample $K$ $Y_i$'s, such that $\sum_{i=1}^{K-1} Y_i < T$ and $\sum_{i=1}^{K} Y_i \geq T$.  Then the sampled item's failure time, $Z$, is calculated to be \[Z = d*\left[(K-1)+\left(T-\sum_{i=1}^{K-1} Y_i\right)/Y_K\right].\]
Notice that $K$ is a random variable. We are defining the distribution $Z$ as the distribution of bootstrapped failure times. Since the degradation is only observed at time increments of $d$, the definition of $Z$ uses a linear interpolation to generate a failure time.   

A quantile can be estimated directly from one bootstrap sample, or multiple samples can be obtained which provides some Monte Carlo uncertainty of the estimate.  Obtaining multiple samples can be computationally intense. Suppose we estimate the distribution of failure times with Monte Carlo samples. The resulting draws can be used to estimate any quantile, however in order to assess uncertainty, the process must be repeated many times to produce multiple draws for the quantile of interest. One issue that makes the Monte Carlo approach less appealing when estimating quantiles is that the Monte Carlo error increases quite dramatically when finding quantiles far from 0.5.  Our proposed procedure to estimate quantiles is much faster and more accurate than the Monte Carlo approach.

To calculate this bootstrapped failure time distribution using the convolutional method  we find the CDF directly.  For a given time $t$ we find a value for $P(Z \leq t)$.  To do this, first find the integer $k$ such that $(k-1) \,  d < t$ and $k \, d\geq t$. Next, define a grid, $\bm{s}$, of length $N$, where $s_0=0$ and $s_{N-1}=k \times \max\{x_i\}$.  Again, $N$ controls the accuracy of this approach, where larger $N$ is preferable. Next, obtain the DFT of $Y$ and $aY$, where $a = t/d-(k-1)$.  Now the approximated DFT sequence of $Z$ is \[ \tilde{f}_{Z,k} = \tilde{f}_{aY,k} \left(\tilde{f}_{Y,k}\right)^{k-1}.\]
The final steps are to invert the DFT sequence of $Z$ and then sum the probability mass on the support grid which is less than $t$.  Quantiles can found by trying different values of $t$. 

\section{Comparisons}

\noindent
One advantage of using the convolutional method is that the computation time can be much shorter when compared to traditional Monte Carlo methods. The time needed to generate bootstrap intervals using the Monte Carlo method is based on the number of bootstrap samples. In contrast, the computation time for the discrete convolutional method is based on the length of $\bm{s}$.  When a small number of bootstrap samples are taken using the Monte Carlo approach, there is little difference in computation time for the two methods. However, in some cases a large number of Monte Carlo samples are required and here the convolutional method will be faster with more accuracy.

\begin{figure}
    \includegraphics[width=80mm]{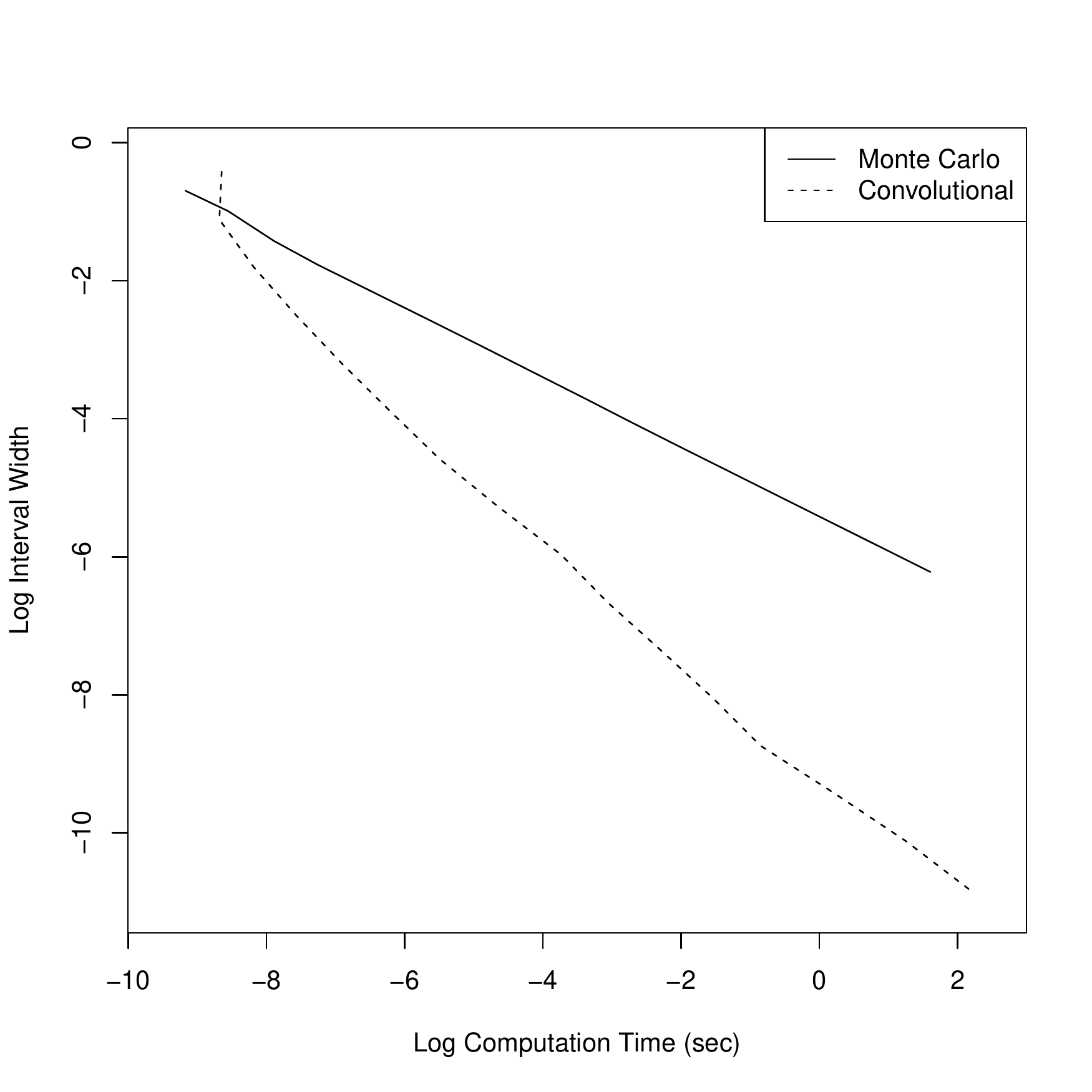}
    \centering
    \caption{A comparison of the two methods shows that the increase in computation time for smaller interval widths is much more dramatic with the Monte Carlo methods.}
    \label{fig:comp}
\end{figure}

The accuracy of the convolutional method is determined by the agreement of $\bm{s}$ and the support of the sample.  In cases where the support of the sample is given to a fixed precision (e.g., rounded), the user defined grid, $\bm{s}$, can often be defined to include all possible bootstrap outcomes. In these instances, increasing the number of support points on the grid will not improve the results (since they are exact).  When a user defined grid cannot include all possible bootstrap samples, mathematical bounds can be placed on the bootstrap distribution. These bounds are not easily comparable to the stochastic bounds from Monte Carlo bootstrap confidence intervals. However, the narrowing of the mathematical bounds occurs at a faster rate than that of the Monte Carlo confidence intervals. This is demonstrated in Figure \ref{fig:comp}, which shows that for wide intervals, the Monte Carlo and discrete convolutional methods are comparable in terms of computation time. However, when more computation time is given, the discrete convolutional method is able to produce narrower intervals than the Monte Carlo bootstrap. Since the figure displays times and interval widths on the log scale, the differences are more obvious. 

Monte Carlo confidence intervals narrow slowly as the computation time is increased.  For the convolutional method the narrowing is approximately linear, in that doubling the computation time shrinks the interval by about one half.  The Monte Carlo algorithm is often simple to implement with fast approximate answers, but when higher precision is desired, it slows considerably. 

Figure \ref{fig:comp} displays the computational time to approximate the bootstrap distribution of the sample mean for 20 observations.
Since there is some natural variation in computation times this simulation was conducted 750 times and averaged together.   
The times for the Monte Carlo intervals are given by the solid curve and the times for the convolutional intervals are given by the dashed curve. 
The curves provide an approximation of the average computation times to achieve a given interval width.
Notice that for the wider intervals the difference in computation time between the two methods is much less pronounced. In fact, when a wide interval is permissible, it is more computationally efficient to use the Monte Carlo bootstrap. However, the Monte Carlo method will only ever be able to generate stochastic bounds, whereas the discrete convolutional method is able to mathematically bound the estimate.

Referring again to Figure \ref{fig:comp}, the convolutional method's computation times generally increase with the number of grid points selected along the support. 
Similarly, the Monte Carlo bootstrap also has narrower intervals as the number of samples (i.e., computation time) increases.  However, the rate at which they narrow are clearly distinct, with the convolutional method converging faster.

Table \ref{tab:times_comp} contains the discrete convolutional method times for the simulations in Figure \ref{fig:comp}. Note that the length of each support grid $\bm{s}$ is a power of two, which takes advantage of this computational efficiency of the FFT. For samples that have high precision, a finer grid is needed to better capture the exact behavior of the bootstrapped distribution. Even with very fine grids the convolutional method is quite fast.

\begin{table}[ht!]
\normalsize
  \begin{center}
    \caption{Average computational times for the convolutional method for varying number of grid points in the support, $\bm{s}$.}
    \label{tab:times_comp}
    \begin{tabular}{c|c}
      \textbf{Support Length} &
      \textbf{Time (sec)} \\
      \hline
    $2^8$ & 0.0003 \\
    $2^{10}$ & 0.0004  \\
    $2^{12}$ & 0.0015 \\
    $2^{14}$ & 0.0055 \\
    $2^{16}$ & 0.0290 \\
    $2^{18}$ & 0.1181 \\
    $2^{20}$ & 0.7438 \\
    $2^{22}$ & 5.2675\\ 
    
   \end{tabular}
  \end{center}
\end{table}

We also compare the convolutional method with saddlepoint approximations. We do this in several examples in the following section.  Generally, the results from the convolutional method seem to be more accurate (i.e., they agree more closely with the Monte Carlo results) than the results from saddlepoint approximations. 

\section{Applications}

A number of data applications have been chosen for inclusion in this article. These examples demonstrate the wide applicability of the convolutional method for bootstrapping. In each of the examples, saddlepoint approximations serve as the primary competitor to the convolutional method, and the Monte Carlo approach (with a very large number of samples) is considered to be the standard.  

We argue that the convolutional approach has some distinct advantages over the saddlepoint method.  First, the saddlepoint method is known to be quite accurate in the tails of distributions, but can be arbitrarily bad at specific points \citep{collins2009nonparametric}.  The advantage of the convolutional method is that it can bound the error of the CDF at any point on the support.  Another advantage is that the convolutional approach uses discrete distributions with compact support---which matches most bootstrap distributions.  In contrast, the saddlepoint approximations are continuous and have infinite support.  The final advantage is that the convolutional method is faster and easier to set up compared with the saddlepoint method, which requires at least calculating a second derivative of the cumulant generating function of the bootstrap distribution.  We demonstrate these advantages in the following examples.

\subsection{Example 2 from \cite{davison1988saddlepoint}}

One example that lends itself well to comparison comes from \cite{davison1988saddlepoint}. In this example, the authors look at the bootstrap distribution of the sample mean for the following 10 numbers:
\begin{table}[ht]
 \normalsize
    \centering
$\bm{x} = $ \{-8.27, -7.46, -4.87, -2.87, -1.27, -0.67, -0.57, 3.93, 6.13, 15.93\}.
\end{table}

Note that these values differ from those reported in the original paper. In the example from \cite{davison1988saddlepoint}, the non-centered data were reported but were centered before performing the bootstrap; the centered observations are what have been reported above.

In the source article, 50,000 Monte Carlo samples were taken to approximate the distribution. With advances in computational capability, more precise approximations have been generated using 200M (million) Monte Carlo samples for this paper. We report the results from the 200M samples. Although Monte Carlo methods have no trouble estimating this distribution, to obtain this precision, it took 20.02 minutes of computation time.

The results using convolutional methods are very similar to the Monte Carlo estimates. However, the time needed for a machine to generate the convolutional results is practically instantaneous at just 0.0673 seconds.  Table \ref{tab:times1} shows the resulting mean absolute error (MAE) and mean squared error (MSE) between the two methods, which vanishes as the number of Monte Carlo samples increases.  This implies the Monte Carlo results are converging to the convolutional results.

\begin{table}[ht!]
 \normalsize
  \begin{center}
    \caption{Davison Example 2 comparisons.  The error between the Monte Carlo and the convolutional bootstraps for varying MC sample sizes.  As the number of Monte Carlo samples increase, both the absolute and squared errors get smaller.  The times reported are for the Monte Carlo results.}
    \label{tab:times1}
    \begin{tabular}{c|c|c|c}
      \textbf{Monte Carlo Samples} &
      \textbf{Time (sec)} &
      \textbf{MAE} &
      \textbf{MSE} \\
      \hline
    100,000 & 0.63 & $3.68 \times 10^{-2}$ & $3.80 \times 10^{-3}$\\
    500,000 & 3.04 & $1.27 \times 10^{-2}$ & $5.84 \times 10^{-4}$\\
    2,500,000 & 14.90 & $7.38 \times 10^{-3}$ & $1.33 \times 10^{-4}$\\
    10,000,000 & 59.55 & $6.19 \times 10^{-3}$ & $1.32 \times 10^{-4}$\\
    200,000,000 & 1,201.38  & $1.25 \times 10^{-3}$ & $2.50 \times 10^{-5}$ \\
   \end{tabular}
  \end{center}
\end{table}

Naturally, the time needed to generate Monte Carlo samples is prone to change with simulation size. Nevertheless, we emphasize that the results from the convolutional method are exact.

\begin{table}[ht!]
 \normalsize
  \begin{center}
    \caption{Comparing the quantile estimates of the three methods using the data from Davison Example 2.  Note that the results of the convolutional approach have been rounded in order to create a fair comparison with the saddlepoint method.}
        \label{tab:table1}
    \begin{tabular}{c|c|c|c}
      \textbf{Probability} &
      \textbf{Monte Carlo} & \textbf{Convolutional} & \textbf{Saddlepoint} \\
      \hline
    0.0001 & -6.31 & -6.31 & -6.31 \\
    0.0005 & -5.78 & -5.78 & -5.78 \\
    0.001 & -5.52 & -5.52 & -5.52 \\
    0.005 & -4.80 & -4.80 & -4.81 \\ 
    0.01 & -4.43 & -4.43 & -4.43 \\
    0.05 & -3.33 & -3.33 & -3.33 \\
    0.10 & -2.69 & -2.69 & -2.69 \\
    0.20 & -1.86 & -1.86 & -1.86 \\
    0.80 & 1.79 & 1.79 & 1.80 \\
    0.90 & 2.85 & 2.85 & 2.85 \\
    0.95 & 3.75 & 3.75 & 3.75 \\
    0.99 & 5.47 & 5.47 & 5.48 \\
    0.995 & 6.13 & 6.13 & 6.12 \\
    0.999 & 7.46 & 7.46 & 7.46 \\
    0.9995 & 8.01 & 8.01 & 7.99 \\
    0.9999 & 9.11 & 9.11 & 9.12 \\
   \end{tabular}
  \end{center}
\end{table}

The saddlepoint approximation suggested by Davison and Hinkley can be seen in Table \ref{tab:table1}. Compared to the Monte Carlo estimates the convolutional approach has an MAE of 0 and the saddlepoint's MAE is 0.004. In general, these approximations are extremely close to the convolutional and Monte Carlo results.  Though the convolutional and Monte Carlo results agree more often as indicated by the MAE. 

Again, the results from the convolutional method described in this application are not an approximation. The distribution of $\bar{Y}$ is discrete and with no need to introduce discretization error, the quantiles described above are exact.  Code for this and the following examples is found in the appendix.

\subsection{Example 3 from \cite{davison1988saddlepoint}}

Davison and Hinkley provide another interesting example in the same paper. This application studies the following twelve matched pairs differences with

\begin{table}[ht]
    \centering
   \normalsize
$\bm{x} = $ \{4.5, -34.2, 7.4, 12.6, -2.5, 1.7, -34.0, 7.3, 15.4, -3.8, 2.9, -4.2\}.
\end{table}

While a bootstrap distribution of a statistic is still of interest, the convolution in question is somewhat different. Every value will be represented in the mean, however the sign has the ability to change.  
In other words, we are looking at the following convolution:
\[ \bar{Y} = \frac{Y_1 + Y_2 + ... + Y_{12}}{12} \]
Where $Y_1$ can take the values -4.5 and 4.5, $Y_2$ can take the values -34.2 and 34.2, and so on. The probability of the positive or negative value appearing in a Monte Carlo sample is 0.5 for each of the random variables.

In our comparisons, we found the distribution for $\bar{Y}$ using six different Monte Carlo sample sizes. For each Monte Carlo sample, the 12 $\times$ 1 ``sign" vector of -1 and 1's was generated, -1 and 1 having the same probability of occurring in each element. The bootstrap sample was calculated as the mean of the element-wise product of the data and sign vectors. The time needed to find 1B (billion) Monte Carlo samples was more than 1.5 hours. 
As with the first example, the Monte Carlo estimates are very accurate. However, in much less time, exact quantiles were found using the convolutional method.  Table \ref{tab:times2} shows that as the number of Monte Carlo samples increases, the approximate Monte Carlo bootstrap distribution converges to the results of the convolutional method.

\begin{table}[ht!]
\normalsize
  \begin{center}
    \caption{Davison Example 3 comparisons.  The error between the Monte Carlo and the convolutional bootstraps for varying MC sample sizes.  As the number of Monte Carlo samples increases, both the absolute and squared errors get smaller.  The times reported are for the Monte Carlo results.}
    \label{tab:times2}
    \begin{tabular}{c|c|c|c}
      \textbf{Monte Carlo Samples} &
      \textbf{Time (sec)} &
      \textbf{MAE} &
      \textbf{MSE} \\
      \hline
    100,000 & 0.66 & $4.39 \times 10^{-4}$ & $4.34 \times 10^{-7}$\\
    500,000 & 3.15 & $2.72 \times 10^{-4}$ & $1.38 \times 10^{-7}$\\
    1,000,000 & 6.22  & $2.28 \times 10^{-4}$ & $1.41 \times 10^{-7}$\\
    10,000,000 & 62.03 & $4.84 \times 10^{-5}$ & $6.22 \times 10^{-9}$\\
    300,000,000 & 1,865.03 & $2.26 \times 10^{-5}$ & $1.02 \times 10^{-9}$ \\
    1,000,000,000 & 6,225.36 & $8.42 \times 10^{-6}$ & $1.56 \times 10^{-10}$\\
    
   \end{tabular}
  \end{center}
\end{table}

Table \ref{tab:table2} compares the results of the three methods.  The results were probabilities from the CDF at values specified in \cite{davison1988saddlepoint}. The results from the convolutional method are exact, however, the values presented in Table \ref{tab:table2} have been rounded to 5 decimal places.  Measuring the distances from the Monte Carlo results we get an MAE of 0.00001 and 0.00364 for the convolutional and saddlepoint methods, respectively.  

\begin{table}[ht!]
\normalsize
  \begin{center}
    \caption{Example 3 from \cite{davison1988saddlepoint}. In general, the results from the convolutional method matched with those from the Monte Carlo bootstrap, but were obtained in a much shorter time. The results from the saddlepoint approximation do not agree as closely.}
    \label{tab:table2}
    \begin{tabular}{c|c|c|c}
      \textbf{Value to Est.} &
      \textbf{Monte Carlo} &
      \textbf{Convolutional} & \textbf{Saddlepoint} \\
      \hline
    Pr($\bar{Y} \leq$ -10.77) & 0.00024 & 0.00024 & 0.00018 \\
    Pr($\bar{Y} \leq$ -10.32) & 0.00097 & 0.00098 & 0.00098 \\
    Pr($\bar{Y} \leq$ -8.97) & 0.01270 & 0.01270 & 0.01300 \\
    Pr($\bar{Y} \leq$ -8.53) & 0.02051 & 0.02051 & 0.02150 \\
    Pr($\bar{Y} \leq$ -7.63) & 0.04420 & 0.04419 & 0.04300 \\
    Pr($\bar{Y} \leq$ -6.28) & 0.09719 & 0.09717 & 0.08790 \\
    Pr($\bar{Y} \leq$ -4.04) & 0.20388 & 0.20386 & 0.20000 \\
    Pr($\bar{Y} \leq$ -2.24) & 0.31106 & 0.31104 & 0.32200 \\
    Pr($\bar{Y} \leq$ -0.90) & 0.41724 & 0.41724 & 0.42700 \\
    Pr($\bar{Y} \leq$ 0.00) & 0.49999 & 0.50000 & 0.50000 \\
    
   \end{tabular}
  \end{center}
\end{table}

In this example, the convolutional method has better agreement with the Monte Carlo estimates than the saddlepoint method. There are a few quantiles whereon both the convolutional and saddlepoint methods agree, however the accuracy of the saddlepoint never exceeds the accuracy of the convolutional method. Once again, the computation time for the convolutional method is quite fast at 0.1130 seconds.

Generating the results from the convolutional method follows the procedure provided in Section \ref{sec:methods} with one major exception. Rather than describing just one random variable, a random variable was needed for each of the pairs. From that point on, the basic theory of using multiplication in the Fourier domain to compute a convolution was utilized as in the previous example.

\subsection{Laser Data Analysis from \cite{balakrishnan2019nonparametric}}

\cite{balakrishnan2019nonparametric}
provide an application using laser degradation data; this analysis lends itself well to comparison.  
Their application looks at cumulative laser device degradation using several different values as failure thresholds.  In their analysis, they  use a saddlepoint method to approximate the distribution of the 90th percentile of the time it takes to attain a cumulative degradation threshold. Here we use the phrase ``failure time" to denote the random time until the designated failure threshold is achieved. Note that in the data, there are 16 degradation measurements per laser for 15 distinct lasers, resulting in a total of 240 measurements.  The data can be found in Table C.17 of \cite{meeker2014statistical}.

In order to find the 90th percentile of failure times, the following Monte Carlo resampling method was used. One of the 15 lasers was randomly chosen. Degradation measurements on the selected laser were randomly sampled (with replacement) until the cumulative degradation surpassed the threshold. Since the lasers were checked in increments of 250 hours, the exact failure times cannot typically be obtained. The final bootstrapped time was determined using linear interpolation on the last degradation measurement and the 250 hours. This process was repeated 40,000 times so that 40,000 bootstrapped failure times were collected. The 90th percentile of the bootstrapped times was recorded. This was then repeated 30,000 times, resulting in an approximation for the bootstrap distribution of the 90th percentile of failure times.

The convolutional method for this analysis also must take into account the specific laser effect, thus we extend the notation from Section \ref{subsec:MethodDeg} to accommodate distinct laser devices.  We denote the $j^{\text{th}}$ measurement from the $i^{\text{th}}$ laser as $x_{i,j}$ (for $j \in \{1,2,\ldots,16\}$ and $i \in \{1,2,\ldots,15\})$.  These observations in turn define 15 discrete probability measures, with random variables $Y_i$, one for each laser device.

To obtain the $i^{\text{th}}$ laser failure time, $Z_i$, the method is the same as in Section \ref{subsec:MethodDeg}.  But accounting for all 15 lasers we must mix these CDFs.  Thus for the bootstrap failure random variable, $Z$, its CDF is
\[ P(Z\leq t) = \frac{1}{15} \sum_{i=1}^{15} P(Z_i\leq t). \]

The distribution of $Z$ combines the information of all lasers, but does not allow the data from one laser to affect the time of another.  To find the 90th percentile of $Z$, a range of times was investigated.

The results of this analysis are included in Table \ref{tab:table3}.  Notice that for each of the selected thresholds, the convolutional estimate is included in the Monte Carlo bootstrap interval. It is also interesting to note that the accuracy of the convolutional method seems to improve as the threshold increases. For a threshold of 1, these Monte Carlo estimates were found in 15.6 hours, however, 4.73 days were needed to find the estimates associated with a threshold of 10. The time required to find all ten estimates (upper and lower bounds) using the convolutional method was less than 6 minutes.

\begin{table}[ht!]
\normalsize
  \begin{center}
          \caption{Laser Data Analysis. Here, (*) indicates that the estimate was not contained in the Monte Carlo bootstrap interval. Notice that each bootstrap interval contains the corresponding quantile, as obtained through the convolutional approach.}
      \label{tab:table3}
    \resizebox{\textwidth}{!}{%
    \begin{tabular}{c|c|c|c|c}
      \textbf{Threshold} &
      \textbf{MC Mean} &
      \textbf{95\% Bootstrap Int} &
      \textbf{Convolutional} &
      \textbf{Saddlepoint} \\
      \hline
    1 & 722.91 & (721.15, 725.18) & 722.82 $\pm$ 0.05 & 724.12 \\
    2 & 1362.20 & (1358.90, 1365.51) & 1362.24 $\pm$ 0.05  & 1365.26 \\
    3 & 2000.17 & (1995.64, 2004.43) & 2000.27 $\pm$ 0.05 & 1997.32 \\
    4 & 2628.74 & (2623.49, 2633.97) & 2628.75 $\pm$ 0.04 & 2626.04 \\
    5 & 3260.88 & (3254.95, 3266.87) & 3260.86 $\pm$ 0.03 & 3253.29* \\
    6 & 3888.47 & (3881.36, 3895.50) & 3888.51 $\pm$ 0.04 & 3879.87* \\
    7 & 4518.72 & (4511.00, 4526.42) & 4518.71 $\pm$ 0.06 & 4506.15* \\
    8 & 5146.46 & (5137.72, 5155.19) & 5146.49 $\pm$ 0.04 & 5132.35* \\
    9 & 5776.05 & (5766.74, 5785.41) & 5776.03 $\pm$ 0.06 & 5758.58* \\
    10 & 6404.09 & (6393.73, 6414.48) & 6404.11 $\pm$ 0.04 & 6384.90* \\
   \end{tabular}
   }
  \end{center}
\end{table}

In this more complicated example, the convolutional method tends to agree with the Monte Carlo samples more often than the saddlepoint method.  When treating the mean Monte Carlo results as the truth, the mean absolute error for the convolutional method is 0.038, while the MAE for the saddlepoint method is 8.935. It appears that the convolutional method has a clear advantage in terms of accuracy.  

\subsection{Numerical Example from \cite{palayangoda2020improved}}

In a more recent article, these laser data were analyzed again. The analysis was similar to that in \cite{balakrishnan2019nonparametric}, however in \cite{palayangoda2020improved}, data from individual lasers were treated as indistinguishable.  More specifically, rather than only sampling from one laser at a time, sampling could occur from any laser degradation measurement.

Formally, the setup for the convolutional method in this example is shown in Section \ref{subsec:MethodDeg}.  Here all the data, $x_{i,j}$ for $i \in \{1,2,\ldots,15\}$ and $j \in \{1,2,\ldots,16\}$, define the discrete random measure for the random variable $Y$.

\begin{table}[ht!]
\normalsize
  \begin{center}
           \caption{Numerical Example from \cite{palayangoda2020improved}. Once again, (*) is an indicator that the saddlepoint approximation was not contained in the Monte Carlo bootstrap interval. For each threshold investigated, the convolutional results are in closer agreement with the Monte Carlo results than the saddlepoint approximation.}
           \label{tab:table4}
    \resizebox{\textwidth}{!}{%
    \begin{tabular}{c|c|c|c|c}
      \textbf{Threshold} &
      \textbf{MC Mean} & 
      \textbf{95\% Bootstrap Int} &
      \textbf{Convolutional} & \textbf{Saddlepoint} \\
      \hline
    1 & 674.0 & (671.4, 676.7) & 674.06 $\pm$ 0.04 & 682.5* \\
    2 & 1248.7 & (1244.7, 1252.3) & 1248.80 $\pm$ 0.04  & 1247.5 \\
    3 & 1792.9 & (1788.7, 1797.1) & 1792.92 $\pm$ 0.03  & 1795.0 \\
    4 & 2330.1 & (2325.0, 2335.2) & 2330.12 $\pm$ 0.03  & 2332.5 \\
    5 & 2862.6 & (2856.8, 2868.4) & 2862.62 $\pm$ 0.03  & 2865.0 \\
    6 & 3391.5 & (3385.1, 3398.1) & 3391.57 $\pm$ 0.03  & 3395.0 \\
    7 & 3917.7 & (3910.5, 3924.9) & 3917.72 $\pm$ 0.03  & 3920.0 \\
    8 & 4441.6 & (4433.8, 4449.4) & 4441.58 $\pm$ 0.03  & 4442.5 \\
    9 & 4963.5 & (4955.3, 4971.9) & 4963.53 $\pm$ 0.03  & 4965.0 \\
    10 & 5483.9 & (5475.0, 5492.8) & 5483.85 $\pm$ 0.03  & 5482.5 \\
    
   \end{tabular}
   }
  \end{center}
\end{table}

In this example, the Monte Carlo estimates were calculated using 30,000 bootstrapped percentiles that were each calculated from 150,000 bootstrapped failure times.  Table \ref{tab:table4} shows the results from the three competing methods.  
When assuming the mean Monte Carlo estimates are the truth, the MAE for the convolutional method is 0.041 while the MAE for the saddlepoint method is 2.62.  As in the previous comparison, the convolutional method appears to have an accuracy advantage over the saddlepoint method presented in \cite{palayangoda2020improved}.

In this simplified analysis, the Monte Carlo bootstrapping method had shorter computation times. It took 3.77 days to find Monte Carlo estimates for the distribution associated with a threshold of 10.  Once again, the computation time required for the convolutional method is substantially shorter than the time needed for the Monte Carlo method, clocking in at just 16 seconds for all 10 thresholds and for both the lower and upper bounds. Although no computation times were reported for the saddlepoint method in \cite{palayangoda2020improved}, we assume they were much faster than the Monte Carlo method.

\subsection{Asthma Application from \cite{warr2020bayesian}}

\begin{figure}[ht]
    \includegraphics[width=80mm]{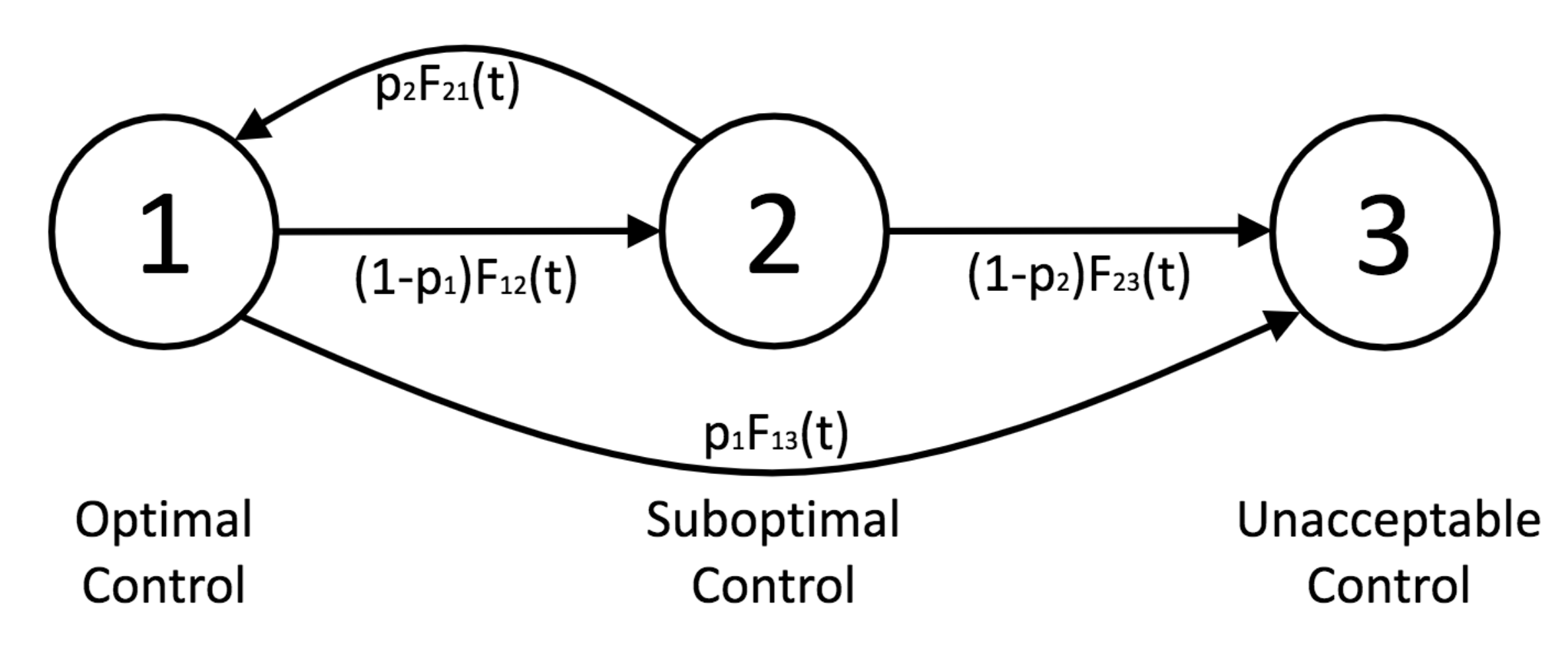}
    \centering
    \caption{Here the states and transitions of the asthma data are depicted. Each transition defines a unique distribution and occurs with fixed probability conditioned on the starting state. The figure has been adapted from \cite{warr2020bayesian}.}
    \label{fig:asthma_diag}
\end{figure}

One final example to consider is found in \cite{warr2020bayesian}. The data for this application is found in the \texttt{SemiMarkov} package in \texttt{R}. The original study was conducted by observing asthma patients in France. The goal of this analysis is to approximate the first passage distribution from stage 1 to stage 3 asthma.  See Figure \ref{fig:asthma_diag} (adapted from \cite{warr2020bayesian}) for a graphical depiction of the states and transitions of asthma patients.

The bootstrapping technique for this example is fairly straightforward. Each bootstrapped patient is assumed to start in stage 1. The next stage was chosen randomly with probabilities determined by sample proportions. A time was then randomly sampled from the observed times for that particular transition. This process was repeated until the bootstrapped patient transitioned into stage 3. In order to produce a comparison, one million Monte Carlo bootstrap samples were generated.

Furthermore, we computed two saddlepoint approximations for comparison. These approximations were also computationally quite fast, however a large degree of smoothing occurred. In order to get more accurate approximations, additional derivatives would need to be computed, making these approximations more difficult to obtain as accuracy increases. 

The first of these two approximations is a second-order saddlepoint method (referred to as Saddlepoint 1). This method uses the approximation developed in \cite{daniels1954saddlepoint}. The result of this method is an approximate pdf and so numerical integration was needed to convert to the CDF. The second saddlepoint method approximated the CDF directly but required an additional derivative to be computed. Saddlepoint 2 uses the approximation developed in \cite{lugannani1980saddle}.

The process of differentiation was made easier through the use of automatic differentiation. Specifically, the \texttt{R} package, \texttt{numDeriv}, was employed \citep{numDeriv}. This package computes accurate first and second order numerical derivatives.
Since much of the difficulty in using saddlepoint approximations comes from the need to differentiate, automatic differentiation seems like an ideal tool to increase the feasibility of saddlepoint approximations using higher order terms to obtain more accuracy.

The analysis conducted in \citeauthor{warr2020bayesian} differs in that the final result is a posterior curve for the CDF of first passage times. For this reason, the results here will not be directly compared to the results from that paper. However, we note that the methodology introduced in this article lends itself well to this application. In this case, the final result is an approximation of the CDF for the bootstrap distribution of first passage times.  In this application the support of the bootstrap first passage distribution is unbounded, which complicates the mathematical bounding of the estimates.  However, our approach still provides accurate estimates without appealing to more complex approaches. 

Let $\tilde{f}_{i \to j,k}$ denote the DFT sequence for the random variable defined by the time needed to transition from state $i$ to state $j$, and $\tilde{g}_{i \to j,k}$ denote the DFT sequence for the random variable defined by the first passage time needed to transition from state $i$ to state $j$.  Then the DFT for the first passage time from state 1 to state 3 is defined as:
\begin{equation} \label{eq:firstpass}
 \tilde{g}_{1 \to 3,k} = p_1\tilde{f}_{1 \to 3,k} + \frac{(1-p_1)\tilde{f}_{1 \to 2,k} \left((1-p_2)\tilde{f}_{2 \to 3,k}+p_1p_2\tilde{f}_{1 \to 3,k}\tilde{f}_{2 \to 1,k} \right)}{1-(1-p_1)p_2\tilde{f}_{1\to 2,k}\tilde{f}_{2\to 1,k}}, 
\end{equation}
equivalent to Equation (5) in \citeauthor{warr2020bayesian}.
The theory for combining the DFTs in this way can be found in \cite{pyke1961markov}.
Here $p_1$ is the probability of transitioning directly to state 3 (instead of state 2), given the current state is state 1.  $p_2$ is the probability of transitioning directly to state 1 (instead of state 3), given the current state is state 2.  

\begin{figure}
\centering
\begin{minipage}{.5\textwidth}
  \centering
  \includegraphics[width=.95\linewidth]{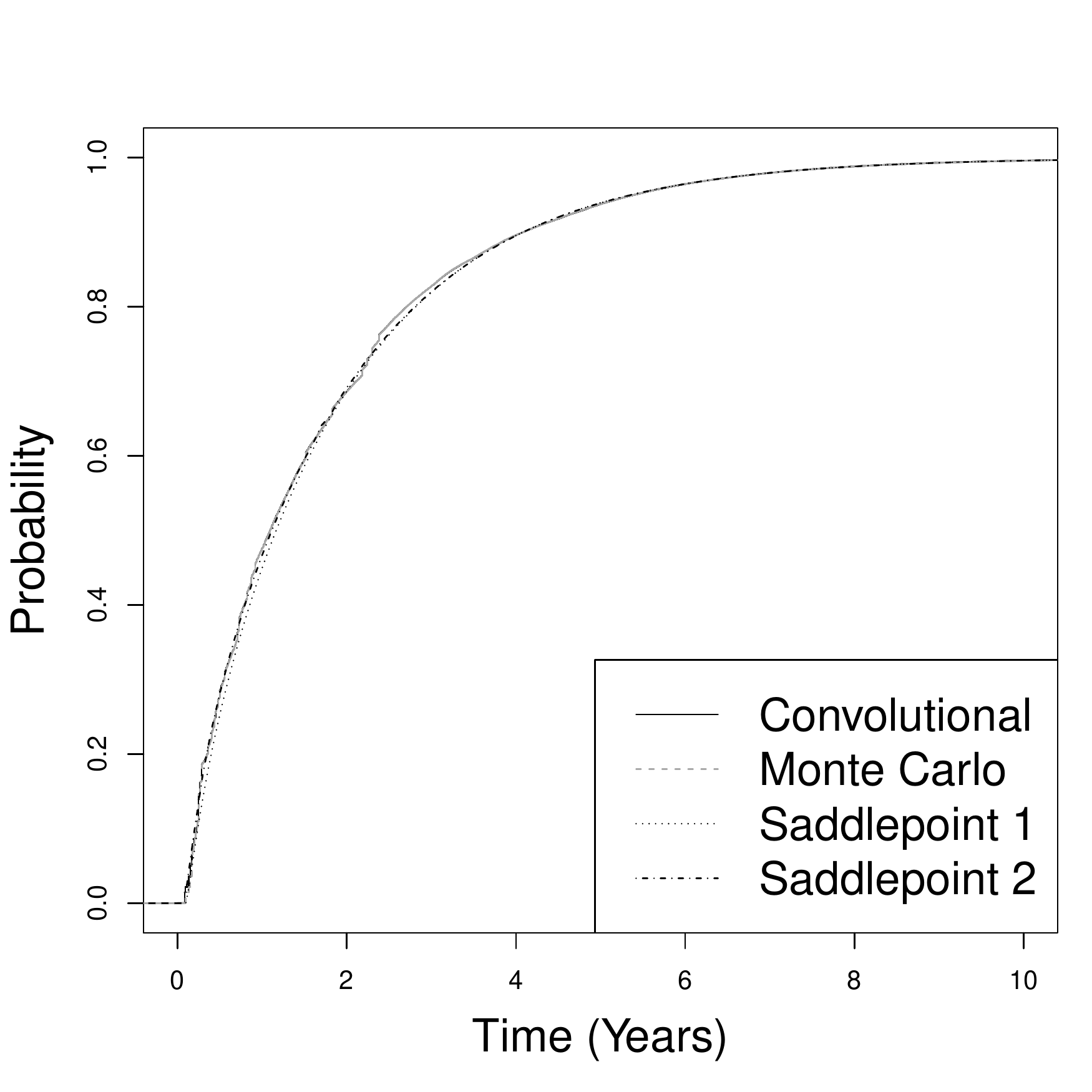}
\end{minipage}%
\begin{minipage}{.5\textwidth}
  \centering
  \includegraphics[width=.95\linewidth]{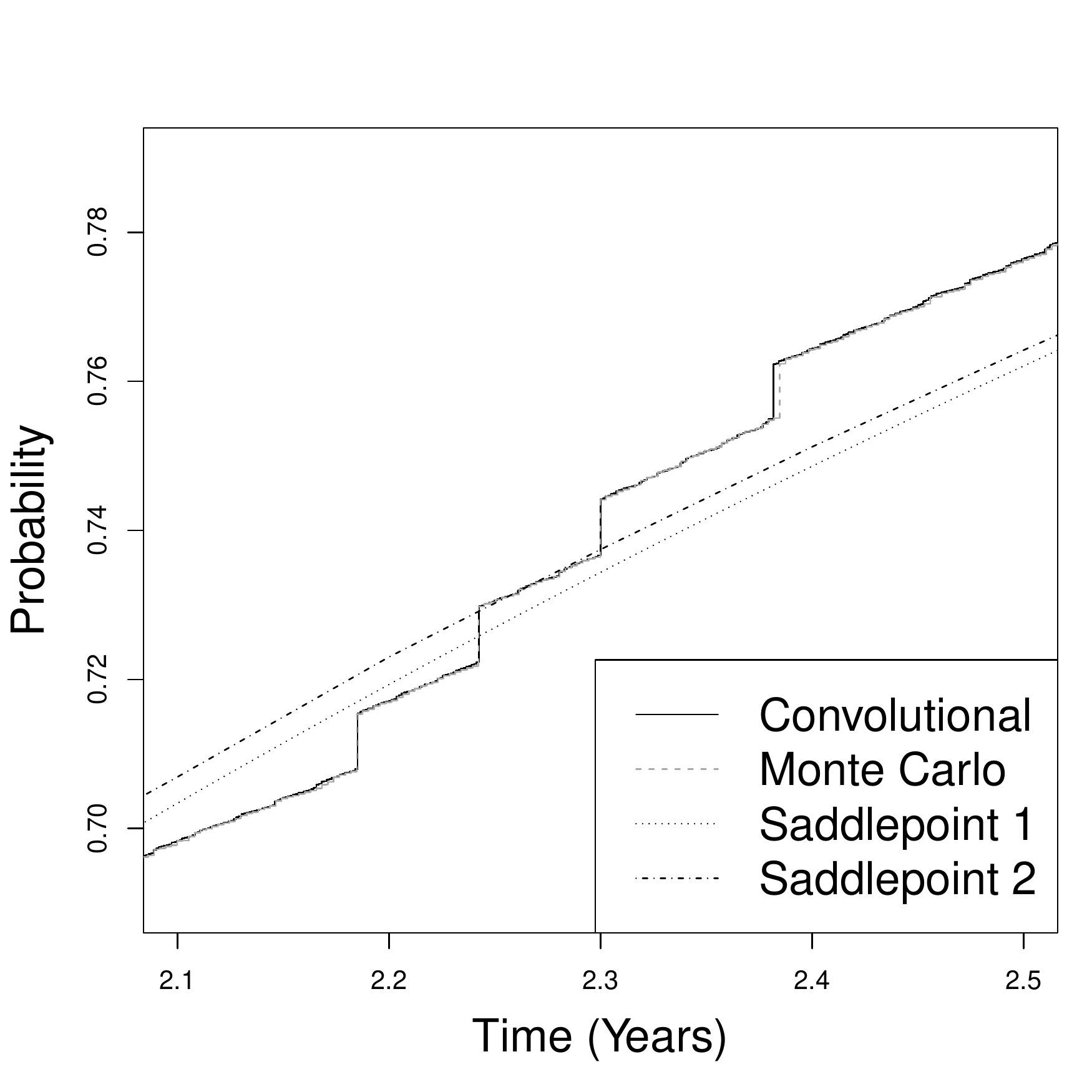}
\end{minipage}
\caption{As seen from the image on the left, the four methods are in fairly close agreement. However, inspecting a close-up view of the approximations in the right image shows that the convolutional and Monte Carlo approaches are able to capture the same local behavior, but the two saddlepoint approaches impose a smoothed continuous approximation to the discrete distribution.}
\label{fig:cdf1}
\end{figure}

Figure \ref{fig:cdf1} displays the bootstrapped first-passage time distribution obtained through the discrete convolutional method. This distribution is compared to the results from the Monte Carlo and saddlepoint methods. Note, from a macro perspective, there is general agreement among the four methods.  However, upon closer inspection it is clear the saddlepoint methods were not able to capture the discrete behavior inherent in this problem.  The MAE between the Monte Carlo and convolutional methods was 0.001, while the MAE between the Monte Carlo and first and second saddlepoint methods was 0.0322 and 0.0216, respectively. Also note that less than 1 second was needed to compute the convolutional result, in contrast to the 11.3 seconds for the first saddlepoint result and 2.54 seconds for the second saddlepoint result. Approximately 6.53 hours of computation time were needed to generate the Monte Carlo samples. In producing the Monte Carlo samples, 100,000 first passage times were produced and then the quantiles of interest were recorded. This was iterated 15,000 times so that the uncertainty could be quantified.

As a point of comparison, a few common quantiles were estimated from the first passage distribution using each method. The results are displayed in Table \ref{tab:asthma}. Again, we point out that the Monte Carlo and convolutional methods agree quite closely. 

\begin{table}[ht!]
\normalsize
  \begin{center}
    \caption{Asthma Application. Quantile estimates using 4 different methods.  The MC Interval is a 95\% Monte Carlo bootstrap interval.  The Monte Carlo and convolutional results are generally quite similar. Due to the smoothed approximation, the saddlepoint results do not agree with the MC results as closely.}
    \label{tab:asthma}
    \resizebox{\textwidth}{!}{
    \begin{tabular}{c|c|c|c|c|c}
      \textbf{Quantile} &
      \textbf{Monte Carlo} & \textbf{MC Interval} &
      \textbf{Convolutional} & \textbf{SP 1} &
      \textbf{SP 2}\\
      \hline
    0.100 & 0.230 & (0.230, 0.233) & 0.229 $\pm$ 0.001 & 0.242  & 0.205 \\
    0.250 & 0.448 & (0.441, 0.449) & 0.448 $\pm$ 0.001 & 0.497  & 0.440 \\
    0.500 & 1.096 & (1.087, 1.112) & 1.094 $\pm$ 0.001 & 1.169  & 1.120 \\
    0.750 & 2.346 & (2.327, 2.366) & 2.344 $\pm$ 0.003 & 2.411  & 2.395 \\
    0.900 & 4.083 & (4.044, 4.120) &4.083 $\pm$ 0.003 & 4.081  & 4.085 \\
   \end{tabular}
   }
  \end{center}
\end{table}

For this application, the CDF was approximately bounded using the theory presented earlier. Table \ref{tab:asthma2} presents the information for bounds generated on different grids. In each case, the grid started at 0 and ended at 30. Even for the grid of size $2^{15}$ the mean width of the error interval was less than 0.0001.
As the grid becomes finer, the average difference in the error bounds appears to approach 0, as seen in Table \ref{tab:asthma2}.  We note that due to the division in Equation \ref{eq:firstpass} the bounds are not guaranteed, although we expect them to be quite accurate.

\begin{table}[ht!]
\normalsize
  \begin{center}
    \caption{Asthma example convolutional error bounds. By increasing the precision of the grid, the upper and lower error bounds for the convolutional method converge.}
    \label{tab:asthma2}
    \begin{tabular}{c|c|c}
      \textbf{Support Length} &
      \textbf{Mean Width} &
      \textbf{Time (s)} \\
      \hline
    $2^{10}$ & 0.00283 & 0.22 \\
    $2^{12}$ & 0.00071 & 0.29 \\
    $2^{14}$ & 0.00018 & 0.65 \\
    $2^{16}$ & 0.00004 & 1.56 \\
   \end{tabular}
  \end{center}
\end{table}

\section{Conclusion}

\noindent
The convolutional bootstrap method provides an attractive alternative to the traditional Monte Carlo and saddlepoint methods. Due to the finite nature of samples, bootstrap distributions are discrete. Relying on this property, it is possible to use discrete computational algorithms to calculate the bootstrap distribution of some statistics.  One simple application of the  discrete convolutional method is in bootstrapping the sample mean. More interesting applications were demonstrated by bootstrapping failure times using degradation data and quantiles of first passage times. 

Additionally, the discrete convolutional method makes it possible to find exact, or at least bounds on quantities of bootstrap distributions. The existence of mathematical error bounds provides a strong case for using the convolutional method.
Even though Monte Carlo bootstrap methods provide confidence intervals, the intervals are stochastic and narrow at a slow rate.

The convolutional bootstrap method is very fast, due to the efficiency of the fast Fourier transform algorithm.  We suppose our method is faster than the saddlepoint approach, and it is much faster than the Monte Carlo approach, for the achieved accuracy.   

In this work we also point out that automatic differentiation could be a powerful tool when using saddlepoint approximations.  
Using this emerging numerical tool 
could allow saddlepoint approximations to be much faster to setup and easier to implement higher order approximations.

As a final note, the convolutional method is limited to statistics that are defined by sums of independent random variables.  For example, we were unable to implement our method for the sample variance, since we know of no way to define it as a sum independent random variables.  
However, we demonstrated the convolutional method to be effective in bootstrapping means, first-passage times, and failure times from degradation data.
The convolutional bootstrap method provides high levels of accuracy with relatively short computation times. 

\subsection*{Acknowledgements}
The authors thank Dave Collins and Lynsie Warr for their content and editorial suggestions.

\subsection*{Conflict of Interest}
The authors have no conflicts of interest to declare that are relevant to the content of this article.
\newpage

\bibliographystyle{dcu}

\newpage
\appendix
\section{Code Appendix}

\subsection{Example 2 from \cite{davison1988saddlepoint}}

\begin{verbatim}
#The centered sample data
x <- c(-8.27,-7.46,-4.87,-2.87,-1.27,-0.67,-0.57,3.93,6.13,15.93)
#Transformed data
w <- round((x-min(x))/length(x),15)
#The support points
s <- (0:25000)/1000;
#Finding the pmf p on the support s
ind <- match(w,s)
p <- rep(0,length(s))
for (i in 1:length(x)) { p[ind[i]] <- p[ind[i]] + 1/length(x) }
#DFT of pmf defined by w 
fp <- fft(p)
#DFT of ybar - min(x)
fm <- fp^length(x)
#pmf of ybar - min(x)
ybar <- Re(fft(fm,inverse=T))/length(s) 
#Probabilities
probs <- c(0.0001,0.0005,0.001,0.005,0.01,0.05,0.1,0.2,0.8,
           0.9,0.95,0.99,0.995, 0.999, 0.9995, 0.9999)
quants <- numeric(length(probs))
for(i in 1:length(probs)){
  quants[i] <- s[min(which(cumsum(ybar)>probs[i]))]+min(x)
}
quants #Results for convolutional method
\end{verbatim}

\subsection{Example 3 from \cite{davison1988saddlepoint}}

See Appendix \ref{verbose} for a step-by-step description of the method and code in this application.

\subsection{Laser Data Analysis from \cite{balakrishnan2019nonparametric}}
    \begin{verbatim}
#Find the data in Meeker and Escobar.
#The data are formatted such that it has 240 rows with the 
# first 16 rows belong to laser-1 etc...
x <- round(read.csv('laserdat.csv')$increase,4)
#Defining constants and the support
N <- 2^18; d <- 250; s <- (0:(N-1))/10000
#Function to calculate CDF of Z using convolutional method
cdf.Z <- function(Z,T){
  m1 <- pmf.y(Z)
  CDF <- function(T) {
    Km1 <- floor(Z/d); index <- max(which(s <= T)); mix <- 0
    for (i in 1:15) { 
      fft.Zi <- fft(m1[i+15,])*fft(m1[i,])^Km1
      mix <- mix + cumsum(Re(fft(fft.Zi,inverse=TRUE))/N)[index] }
    1-mix/15   }
  CDF(T) }
#Function to calculate pmf of y for some value of Z.
# Each row of the output matrix is the pmf of y for the i-th laser
# the last 15 rows includes a linear interpolation of the same 
# pmfs for the degradation in an interval less than 250 hours 
pmf.y <- function(Z) {
  pmf.yi <- matrix(0, nrow=30, ncol=N)
  for (i in 1:15) { pmf.yi[c(i,i+15),] <- yi.indices(i,Z) }
  pmf.yi
}
#Function which returns the indices (of the support) for the 
# point masses of y for the L-th laser.  The first row of the
# matrix is for degradations measured in 250 hour increments.
# The 2nd row is a linear interpolation for smaller increments.
# The last argument determines a lower or upper bound
yi.indices <- function(L,Z){
  prob250 <- numeric(N); probLI <- numeric(N); i <- 1:16
  ind250 <- ceiling(x[i+16*(L-1)]*10000)+1
  num250 <- table(ind250); prob250[unique(ind250)] <- num250/16
  if (lower.bound) {
    indLI <- ceiling(x[i+16*(L-1)]*10000*(Z/d-floor(Z/d)))+1
  } else {
    indLI <- floor(x[i+16*(L-1)]*10000*(Z/d-floor(Z/d)))+1
  }
  numLI <- table(indLI); probLI[unique(indLI)] <- numLI/16
  rbind(prob250, probLI)
}
#Defining a function such that the 90th percentile will be the root
percent90 <- function(z,T) { cdf.Z(z,T)-0.9 }
#0.9 quantiles for thresholds of 1,2,...,10 
lower.quants <- rep(NA,10); upper.quants <- rep(NA,10);
for (i in 1:10) {
  lower.bound=TRUE
  lower.quants[i] <- uniroot(percent90,c(250,6500),T=i)$root
  lower.bound=FALSE
  upper.quants[i] <- uniroot(percent90,c(250,6500),T=i)$root
}
round((lower.quants+upper.quants)/2,2)
\end{verbatim}
    
\subsection{Numerical Example from \cite{palayangoda2020improved}}
\begin{verbatim}
# Run the code from the previous example first!
#Function to calculate CDF of Z using convolutional method
cdf.Z <- function(Z,T){
  m1 <- pmf.y(Z)
  CDF <- function(T) {
    Km1 <- floor(Z/d); index <- max(which(s <= T))
    fft.Z <- fft(m1[2,])*fft(m1[1,])^Km1
    1-cumsum(Re(fft(fft.Z,inverse=TRUE))/N)[index]  }
  CDF(T) }
#Function to calculate pmf of y for some value of Z.
# The 1st row of the output matrix is the pmf of y 
# The 2nd row includes a linear interpolation of the same 
# pmf for the degradation in an interval less than 250 hours 
pmf.y <- function(Z){
  prob250 <- numeric(2^18); probLI <- numeric(2^18)
  for(laser in 1:15){
    for(i in 1:16){
      ind250 <-ceiling(x[i+16*(laser-1)]*10000)+1
      prob250[ind250] <- prob250[ind250] + 1/16
      if (lower.bound) {
        indLI <-ceiling(x[i+16*(laser-1)]*10000*(Z/d-floor(Z/d)))+1
      } else {
        indLI <-floor(x[i+16*(laser-1)]*10000*(Z/d-floor(Z/d)))+1
      }
      probLI[indLI] <- probLI[indLI] + 1/16 } }
  rbind(prob250, probLI)/15
}
#0.9 quantiles for thresholds of 1,2,...,10 
for (i in 1:10) {
  lower.bound=TRUE
  lower.quants[i] <- uniroot(percent90,c(250,6500),T=i)$root
  lower.bound=FALSE
  upper.quants[i] <- uniroot(percent90,c(250,6500),T=i)$root
}
round((lower.quants+upper.quants)/2,2)
\end{verbatim}

\subsection{Asthma Application from \cite{warr2020bayesian}}

\begin{verbatim}
#The data is contained in the SemiMarkov library
library(SemiMarkov); data(asthma)
#Initializing vectors and constants
t12 <- numeric(); t13 <- numeric(); t21 <- numeric(); 
t23 <- numeric(); j <- 0; k <- 0; m <- 0; n <- 0
#Creating separate vectors for each of the transitions
for(i in 1:nrow(asthma)){
  if(asthma$state.h[i] == 1){
    if(asthma$state.j[i] == 2){
      j <- j+1; t12[j] <- asthma$time[i] } 
    if(asthma$state.j[i] == 3){ 
      k <- k+1; t13[k] <- asthma$time[i] }
  }
  if(asthma$state.h[i] == 2){
    if(asthma$state.j[i] == 1){
      m <- m+1; t21[m] <- asthma$time[i] } 
    if(asthma$state.j[i] == 3){
      n <- n+1; t23[n] <- asthma$time[i] }
  } }
#The sample proportions serve as probability estimates
p12hat <- j/(j+k); p13hat <- k/(j+k)
p21hat <- m/(m+n); p23hat <- n/(m+n)
#An upper limit of 20 should capture virtually all the probability
s <- seq(0, 30, length.out=2^15)
#Defining the vectors of probabilities
pmf.yij <- function(times, s){
  probs <- rep(0,length(s)); indices <- rep(0,length(times))
  for(i in 1:length(times)){
    indices[i] <- max(which(s<=times[i])) + upper.bound }
  for (i in 1:length(indices)) {
    probs[indices[i]] <- probs[indices[i]] + 1/length(times) }
  probs
}
upper.bound=TRUE
pmf.12 <- pmf.yij(t12, s); pmf.13 <- pmf.yij(t13, s)
pmf.21 <- pmf.yij(t21, s); pmf.23 <- pmf.yij(t23, s)
#Employing the fft for each transition
ft.pmf.12 <- fft(p12hat*pmf.12); ft.pmf.13 <- fft(p13hat*pmf.13)
ft.pmf.21 <- fft(p21hat*pmf.21); ft.pmf.23 <- fft(p23hat*pmf.23)
#Convolution defined by transition from state 1 -> 3
fft.g13 <- ft.pmf.13+(ft.pmf.12*(ft.pmf.23+ft.pmf.13*ft.pmf.21)/
           (1-ft.pmf.12*ft.pmf.21))
cdf.g13 <- cumsum(Re(fft(fft.g13,inverse=T))/length(s))
plot(s,cdf.g13,type="l") #Plot of the cdf
#Finding quantiles
qCon <- function(prob) s[max(which(cdf.g13<=prob))]
upper.quants <- c(qCon(.1),qCon(.25),qCon(.5),qCon(.75),qCon(.9))
#Repeating for the lower bounds
upper.bound=FALSE
pmf.12 <- pmf.yij(t12, s); pmf.13 <- pmf.yij(t13, s)
pmf.21 <- pmf.yij(t21, s); pmf.23 <- pmf.yij(t23, s)
ft.pmf.12 <- fft(p12hat*pmf.12); ft.pmf.13 <- fft(p13hat*pmf.13)
ft.pmf.21 <- fft(p21hat*pmf.21); ft.pmf.23 <- fft(p23hat*pmf.23)
fft.g13 <- ft.pmf.13+(ft.pmf.12*(ft.pmf.23+ft.pmf.13*ft.pmf.21)/
           (1-ft.pmf.12*ft.pmf.21))
cdf.g13 <- cumsum(Re(fft(fft.g13,inverse=T))/length(s))
#Finding quantiles
lower.quants <- c(qCon(.1),qCon(.25),qCon(.5),qCon(.75),qCon(.9))
(upper.quants+lower.quants)/2
\end{verbatim}

\section{Example 3 from \cite{davison1988saddlepoint}: Verbose Instructions} \label{verbose}

This section is meant to provide a step-by-step example of problem-solving with the convolutional method. The outline assumes that \texttt{R} is being used; however, this outline could be adjusted and applied to other programming languages.

\BeforeBeginEnvironment{verbatim}{\def\baselinestretch{1}}

\begin{itemize}
    \item Start by storing the raw data in a vector.
\begin{verbatim}
x <- c(4.5,-34.2,7.4,12.6,-2.5,1.7,-34.0,7.3,15.4,-3.8,2.9,-4.2)
    \end{verbatim}
    \item Identify the convolution of interest. The convolution of interest for this application is the mean of 12 random variables, each with point mass of probability 0.5 at $x_i$ and $-x_i$, for $i \in \{1,\ldots,12\}$.
    \item Create a separate vector for each of the random variables to be convolved. Each vector should contain the possible realizations for that RV. In this application, the vectors were started all at once by combining the data vectors, \texttt{x}, and \texttt{-x} to create a matrix with two columns.
    \begin{verbatim}
n <- length(x); y <- cbind(x, -x);
    \end{verbatim}
    \item Create a vector that contains the support grid, \textbf{s}. Here the support was shifted to the right since fully positive supports are generally easier to work with. The support from 0 to 70 will allow for probability between -35 and 35.
    \begin{verbatim}
s <- round(seq(0, 70, by=1/120),14)
    \end{verbatim}
    \item At this point multiple empty vectors were created that will be used in the following steps.
    \begin{verbatim}
#Defining some data structures
p <- matrix(0, 12, length(s)); ind <- matrix(NA, 12, 2)
ind2 <- rep(NA,10); fft.p <- matrix(NA, 12, length(s)) 
    \end{verbatim}
    \item Before moving on, we want to make sure that the RVs described by the matrix \texttt{y} are the random variables that will be convolved. Here we shift the data so that it matches the support and then we divide by \texttt{n}. We also round at 14 decimal places to make sure the support numerically matches the adjusted data.
    \begin{verbatim}
#The shifted and scaled data vectors
w <- round((y+35)/12,14)
    \end{verbatim}
    \item For each random variable described by the matrix \texttt{w}, we need to find which support points should have positive probability. In this step we are defining the pmf for each random variable and inputting it as a row in the matrix \texttt{p}.
    \begin{verbatim}
#Defining the pmf of each w on s
for (i in 1:n) { ind[i,] <- match(w[i,],s) }
for (j in 1:n) { for (i in 1:2){
  p[j,ind[j,i]] <- p[j,ind[j,i]] + 1/2
} } 
    \end{verbatim}
    \item Now the FFT is applied to each random variable by using the \texttt{fft} function on each row of the matrix, \texttt{p}.
    \begin{verbatim}
#Computing the fft for each pmf
for (i in 1:nrow(p)) { fft.p[i,] <- fft(p[i,]) }  
    \end{verbatim}
    \item At this point, we perform the convolution of interest using the multiplicative property of the DFT.
    \begin{verbatim}
#Finding the fft for the pmf of ybar
fft.ybar <- apply(fft.p,2,prod)      
    \end{verbatim}
    \item Finally, we are able to invert the FFT to find the pmf for the convolution of interest. Notice that in \texttt{R} we must divide by the size of the support in order to normalize the pmf. We are also finding the cumulative sum in order to obtain the CDF. The quantiles were obtained in order to compare to the results from the source article.
    \begin{verbatim}
#The CDF of ybar
cdf.ybar <- cumsum(Re(fft(fft.ybar,inverse=T))/length(s))
#Finding the CDF values at specific support points
quants <- c(-10.77,-10.32,-8.97,-8.53,-7.63,-6.28,-4.04,-2.24,-.9,0)
for (i in 1:10) { ind2[i] <- max(which(s-35 <= quants[i])) }
cdf.ybar[ind2]
    \end{verbatim}

\end{itemize}

\end{document}